# Robust All-Optical Single-Shot Readout of Nitrogen-Vacancy Centers in Diamond

Dominik M. Irber[1,2], Francesco Poggiali[1,2], Fei Kong[3], Michael Kieschnick[4], Tobias Lühmann[4], Damian Kwiatkowski[5], Jan Meijer[4], Jiangfeng Du[3], Fazhan Shi[3], Friedemann Reinhard[1,2,6*]

[1] TU München, Walter Schottky Institut and Physik-Department, Am Coulombwall 4, 85748 München, Germany
[2] Munich Center for Quantum Science and Technology (MCQST), Schellingstraße 4, 80799 München, Germany
[3] CAS Key Laboratory of Microscale Magnetic Resonance & Department of Modern Physics, University of Science and Technology of China, Hefei 230026, China
[4] Applied Quantum Systems, Felix-Bloch Institute for Solid-State Physics, University Leipzig, Linnéstraße 5, 04103 Leipzig, Germany
[5] Institute of Physics, Polish Academy of Sciences, al. Lotników 32/46, 02-668 Warsaw, Poland
[6] Institut für Physik, Universität Rostock, Albert-Einstein-Str 23, 18059 Rostock, Germany

High-fidelity projective readout of a qubit's state in a single experimental repetition is a prerequisite for various quantum protocols of sensing and computing. Achieving single-shot readout is challenging for solid-state qubits. For Nitrogen-Vacancy (NV) centers in diamond, it has been realized using nuclear memories or resonant excitation at cryogenic temperature. All of these existing approaches have stringent experimental demands. In particular, they require a high efficiency of photon collection, such as immersion optics or all-diamond micro-optics. For some of the most relevant applications, such as shallow implanted NV centers in a cryogenic environment, these tools are unavailable. Here we demonstrate an all-optical spin readout scheme that achieves single-shot fidelity even if photon collection is poor (delivering less than $10^3$ clicks/second). The scheme is based on spin-dependent resonant excitation at cryogenic temperature combined with spin-to-charge conversion, mapping the fragile electron spin states to the stable charge states. We prove this technique to work on shallow implanted NV centers, as they are required for sensing and scalable NV-based quantum registers.

Much of the popularity of NV⁻ centers in diamond is owing to the fact that the readout of its electron spin is straightforward, since fluorescence intensity correlates with the spin state[1]. However, this simple approach is highly inefficient because the relative contrast between the spin states is short-lived (approx. 250 ns) and low (about 30 %)[2]. This corresponds to a single-shot signal-to-noise ratio (SNR) of 0.05 (0.03) for a count-rate of 200 kcps (50 kcps). Thus, averaging over several hundred to several ten thousand of experimental repetitions is necessary to read out the spin state with an SNR of 1.
One option to increase the single-shot SNR is spin-to-charge conversion (SCC)[3,4]. This readout approach maps the fragile spin state to the more robust charge state of the NV center, which can be read out optically with close to 100 % fidelity even at room temperature[5]. This mapping is typically achieved by first shelving the spin $m_s = |\pm 1\rangle$ population to the meta-stable singlet state of the NV⁻ center and subsequently ionizing the NV⁻ center out of the triplet[3] or the singlet[4] state during the lifetime of the latter. So far, SCC has reached readout fidelities of up to 67%, limited by non-deterministic shelving to and storing in the singlet state.
More sophisticated schemes for spin readout have achieved single-shot readout, i.e. a single-shot SNR > 1. A first method exploits repetitive readout from a nearby nuclear ancilla qubit[6]. This method requires a strong and carefully aligned magnetic field, and efficient photon collection for readout to succeed within the lifetime of the nuclear qubit.
A second scheme consists in tuning a narrow-linewidth laser in resonance to a cycling transition in the low-temperature excitation spectrum of the NV⁻ center[7]. In this configuration, the NV⁻ is spin-selectively excited and thus producing fluorescence only if its spin state matches the used optical transition. This gives a high contrast signal for a finite time, limited by spin depolarization due to laser illumination. Therefore, the scheme requires all-diamond micro-optics for efficient photon acquisition. This is in particular not available for NV centers close to a planar sample surface.
Here, we present a single-shot readout scheme that eliminates the need for sophisticated optics. The key of our approach is spin-to-charge conversion at cryogenic temperature, where resonant excitation enables both high spin-selectivity of SCC and efficient readout of the charge state by poor collection optics. In detail, our protocol employs resonant excitation[8] to only excite the NV⁻ if its spin state matches the used optical transition, typically a spin $|0\rangle$ transition. Simultaneous illumination with a high-power 642 nm laser ionizes the NV⁻ from the excited state, while not causing internal excitation dynamics (Fig. 1a). Doing so, we lift the fidelity of SCC well above a single-shot SNR of 1 for a natural NV center microns deep in the diamond ('deep NV'), and to the single-shot threshold for a shallow-implanted NV center closer than 100 nm to the diamond surface. The technique also promises to be robust against strong misaligned background magnetic fields.

**Results**

All measurements were performed in a homebuilt confocal microscope. The sample is in a Helium flow cryostat and can be illuminated through an air objective with numerical aperture of 0.95 simultaneously with three independently gateable lasers: a narrow-band red laser tuned to a strong cycling transition starting from spin state $|0\rangle$ ('resonant laser'); a strong red diode laser for photoionization ('ionization laser'); and a green diode laser for initialization of the charge and (in some experiments) the spin state (Fig. 1b). Besides, the NV⁻ center can be excited by two gateable microwave (MW) sources, tuned to both directly allowed spin transitions within the ground state. A static magnetic field of ~1 mT was applied, which was not aligned along the NV axis (Supplementary Section S.1).
These tools implement the final protocol (Fig. 1c). Its most crucial components are spin readout by (1) a highly spin-selective photoionization step ('spin-dep. ionization') implemented jointly by the resonant and the ionization laser and (2) low-power detection of the charge state by the resonant laser ('readout'), which is made agnostic to the spin state by a strong simultaneous microwave drive ($T_{Rabi} = ~1$ µs). Initialization of the charge state ('charge init.') is performed by the green laser and confirmed by a spin-agnostic probe for later postselection. The spin state is initialized in $|+1\rangle$ by repeated resonant depletion of state $|0\rangle$ followed by emptying of the $|-1\rangle$ state by a microwave pulse.
We first demonstrate the protocol on a deep natural NV center. At cryogenic temperatures line narrowing allows different spin states to be individually addressed[9] and the NV⁻ excited state reveals 6 sublevels. The measured spectrum (Fig. 1d) is well described by the model of Doherty *et al.*[10] (Fig. 1d upper panel) with a non-axial strain of 1.73 GHz. See Supplementary Section S.2 for more details. Two of them have $S_z$ character[8], thus having allowed cycling transitions from ground state spin $|0\rangle$, one of which (-7 GHz) serves as working transition for the red resonant laser. Note that $|0\rangle$ transitions are observed in the $|+1\rangle$ trace because of imperfect spin initialization in combination with the added MW π-pulse.

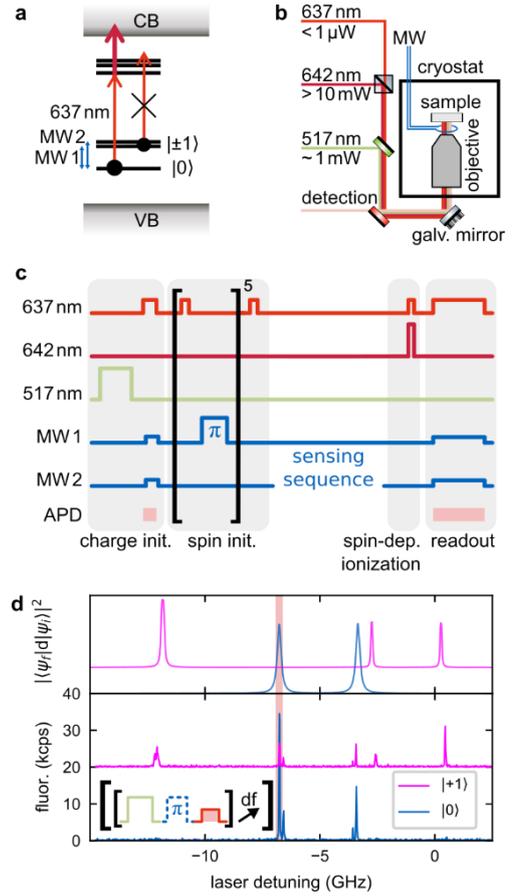

Fig. 1: **Main idea of the readout scheme. a** Energy levels (simplified) of an NV⁻ center in diamond. Initially, the NV⁻ center is in its (optical) ground state with spin state either $|0\rangle$ or $|\pm1\rangle$. If the spin state is $|0\rangle$, a gated laser tuned to a $|0\rangle$ transition (637 nm; thin light red arrow) can excite the NV⁻ center, while spin $|1\rangle$ is protected against excitation. A second gated high-power laser (642 nm; bold dark red arrow) can ionize from the excited state. The blue arrows indicate the microwave (MW) transitions used below. **b** Schematic of the setup. Three individually gated lasers can illuminate the sample, which is mounted in a flow cryostat. In addition, the sample can be driven by two MW frequencies. **c** Final pulse sequence. Red, dark red and green correspond to gated 637 nm resonant, 642 nm and 517 nm laser excitation. Blue refers to MW drive. During photon acquisition ("APD"; rose shade) for postselection and final data acquisition, cw MW excitation at both ground state transitions is added to constantly mix the spin state during charge-state readout. **d** Lower panel: Photoluminescence excitation (PLE) spectrum of the deep NV⁻ center that is also used for Fig. 2 and Fig. 3. Detuning is denoted from 637.20 nm. Off-axial strain is estimated to be 1.7 GHz. The inset shows the used pulsed sequence; 'df' indicates a change of the laser detuning. Upper panel: Simulated spectrum according to Doherty et al.[10] Red highlighting indicates the transition used below.

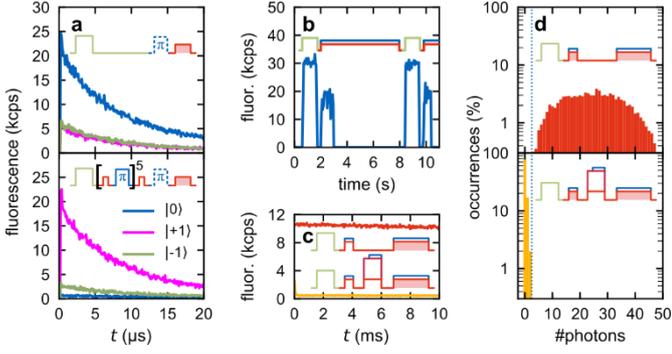

Fig. 2: **Spin and charge state stability of the deep NV center. a** Average fluorescence under optical pumping with a 637 nm laser being resonant to the main $|0\rangle$ transition. In the upper panel, the NV center was initialized to $|0\rangle$ with (70±1)% probability by a simple green laser pulse, followed by no MW excitation (blue) or by a π-pulse on either the $|0\rangle \leftrightarrow |+1\rangle$ or $|0\rangle \leftrightarrow |-1\rangle$ transition (magenta and green curve). The lower panel displays the same measurement after initializing to $|+1\rangle$ with (88±2)% probability using the explicit spin initialization protocol presented in Fig. 1c. The insets are sketches of the sequence used for the upper and lower panel. **b** Charge state stability under excitation with the 637 nm laser plus cw MW excitation at both ground state MW transitions. This was alternated with 1 s of green repumping. This data is not averaged, but just one repetition. **c** Average fluorescence of the NV being preferentially in the charge states NV$^-$ (upper curve) or NV$^0$ (lower curve). This data (without using postselection) was taken simultaneously with the data presented in part d. **d** Distribution of the numbers of fluorescence photons that were detected during 1 ms of readout. The upper panel displays the distribution directly after the charge-state initialization to NV$^-$, as shown in Fig. 1c, while the lower part includes a strong ionization pulse in between initialization and readout for conversion to NV$^0$. The dotted line indicates the threshold used for analysis. Note the logarithmic y scale.

**Spin-state stability under optical pumping and spin-state initialization.** Driving the spin in state $|0\rangle$ on this selected transition with 56 nW (0.08 $P_{sat}$; see Supplementary Fig. S3 for saturation curves) induced fluorescence, which decayed to almost zero within 20 μs (Fig. 2a, upper panel, dark blue curve), as the spin is pumped from spin $|0\rangle$ to $|\pm1\rangle$ due to spin mixing[7]. During these 20 μs, we collected 0.17 photons on average. This low number compared to Robledo *et al.*[7] is due to the fact that we do not use any photonic structures, and precludes direct single-shot readout of the spin by resonant excitation[7] (fidelity of 52.8%; see Supplementary Fig. S.5 and Discussion S.4). This highly spin-selective fluorescence still enables benchmarking of the spin initialization. Using off-resonant excitation by a green laser for simultaneous charge and spin initialization, we obtain a mixture of $|0\rangle:|+1\rangle:|-1\rangle$ = (70±1):(13±1):(16±1) percent, consistent with previous reports[2,11]. The most effective way to improve spin initialization is to pump on an optical spin $|\pm1\rangle$ transition. We decided for an experimentally simpler method; repeating optical depletion of the $|0\rangle$ transition and a π-pulse on the $|-1\rangle$ MW transition, which prepares the spin state $|+1\rangle$ with improved purity ( $|0\rangle:|+1\rangle:|-1\rangle$ = (0±1):(88±2):(12±2) percent; Fig. 2a lower panel). See Supplementary Section S.5 for more details on the spin initialization.

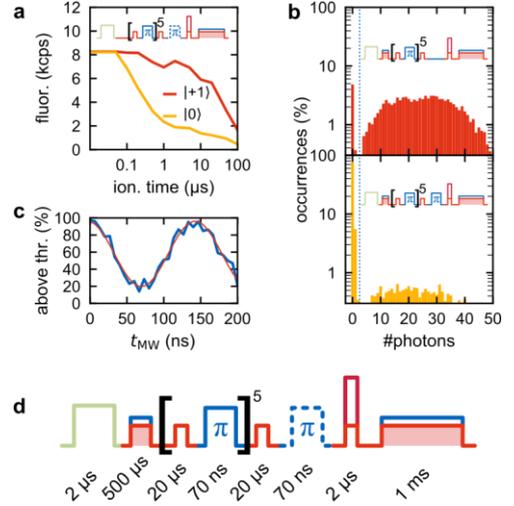

Fig. 3: **Spin-dependent ionization. a** Average fluorescence of the charge-state readout after charge initialization (without postselection), spin initialization according to Fig. 1c and spin dependent ionization with varying time. For parts b and c, 2 μs was used. **b** Distribution of the number of fluorescence photons that were detected during 1 ms of readout, after preparing the NV$^-$ in either spin $|+1\rangle$ or $|0\rangle$ and spin-dependently ionizing it for 2 μs. The as-measured fidelity is (88.5±0.5)%. The dotted line indicates the threshold used for analysis. Sketches of the used sequences are inset. **c** Rabi oscillation measured accordingly to part b. The y-axis is the fraction of experimental repetitions with detected photon number during readout above threshold. The data was measured within about 11 s. The red line is a cosine fit. **d** Sequence as used for part b. The color-code is explained in Fig. 1.

**Charge-state stability and readout.** We also use the resonant laser to read out the charge state[12,13], which is in contrast to most SCC publications so far, which used an orange laser for that purpose. We counteract spin depletion by simultaneously applying cw MW excitation at both ground state MW transitions to constantly mix spin population and in turn re-establish some population in $|0\rangle$. The charge state is stable under this combined excitation. Pumping on the transition with 13 nW (<0.02 $P_{sat}$) plus cw MW, the NV$^-$ gets ionized after a second timescale (Fig. 2b). This can be seen as a sudden decrease in count rate to almost zero, because NV$^0$ has a higher-energy separation between ground and excited state, and is in turn protected against excitation by 637 nm. With the final readout power (56 nW, 0.08 $P_{sat}$), the charge state is stable for more than 10 ms (Fig. 2c). The photon statistics during a 1 ms readout pulse is presented in Fig. 2d. It displays clearly separated distributions for NV$^-$ and NV$^0$ events. In the final readout, events with ≥3 clicks were assigned to be NV$^-$. NV$^-$ events have been produced by initialization using the green laser at 1.4 mW for 2 μs, initializing into the negative NV$^-$ charge state in (46±1)% of repetitions. Postselecting (≥6 clicks within 500 μs; keeping 37% of repetitions) on the charge state after the green illumination, as shown in Fig. 1c, improves initialization to NV$^-$ to (99.7±0.7)%. Postselection also removes repetitions with severe spectral diffusion. NV$^0$ events are produced by first initializing NV$^-$ as described, and appending a 20 μs long ionization pulse of 637 nm plus 642 nm, with cw MW added after 5 μs. Importantly, the high stability of the charge state under resonant excitation enables charge readout with near-perfect ((98.1±0.5)%) fidelity using inefficient collection optics.

**Spin-dependent ionization.** The heart of our readout protocol is the spin-dependent ionization, which is a two-photon process. The second photon is provided by the strong (17 mW) red laser, red-detuned (642 nm) against the NV$^-$ zero phonon line (ZPL). It ionizes from the $^3E$ excited state on a fast (1 μs) time scale, but its energy is by itself insufficient to drive excitation into the $^3E$ state. Besides, it causes negligible stimulated emission back into the $^3A_2$ state[14]. NIR lasers fulfill these criteria, too[13], however we observed much less efficient ionization at 980 nm (see Supplementary Fig. S.9 and Discussion). Ionization is made spin-selective by simultaneously applying the resonant laser, which provides the first photon for excitation into the $^3E$ state. As this laser only excites spin $|0\rangle$, the spin population in the excited state and, hence, ionization is highly deterministic for spectrally well-separated transitions. Fig. 3a shows the averaged fluorescence for charge state readouts after the NV⁻ center has been prepared in spin $|0\rangle$ or spin $|+1\rangle$ and spin-selectively ionized. Fluorescence is higher in the latter case, because spin $|+1\rangle$ is protected against resonant excitation and in turn against ionization. We expect the decay for spin $|+1\rangle$ for long ionization times to stem from residual excitation by the 642 nm laser. We optimized the ionization time for highest fluorescence contrast between the two preparations (resulting in 2 μs; contrast 7.5 kcps vs 1.9 kcps). The resonant power was kept at 56 nm, as optimized for the readout. Fig. 3b is the statistics of photon counts for the whole initialization-ionization-readout protocol (Fig. 3d) for both spin preparations. The measured end-to-end fidelity of the scheme is $(88.5\pm0.5)\%$ (SNR 1.74±0.06). Correcting for imperfect spin initialization (fidelity of $(94.0\pm0.9)\%$) and error of the MW π-pulse, the readout fidelity (comprising only the ionization and charge detection steps) is $(96.4\pm2.2)\%$, which corresponds to a single-shot SNR of 3.5±1.2 (see Supplementary Section S.8). Reducing the readout time for the final charge state from 1 ms to 100 μs, the end-to-end fidelity is only slightly degraded from $(88.5\pm0.5)\%$ to $(82.3\pm0.5)\%$. Importantly, the same performance could be achieved under strongly reduced photon flux (0.5 kcps instead of 50 kcps), if a 10 ms readout window is used (see Supplementary Section S.9). For a long (>10 ms) sensing sequence, this affords a speedup of $10^3$ (for 50 kcps) over standard readout and a factor of 20 over resonant excitation readout[7] (see Supplementary Section S.4). For a short sequence in typical (50 kcps) conditions, our method is still as fast as the standard technique (Supplementary Section S.4). As an example, Fig. 3c shows Rabi nutations measured with 144 repetitions in 11 s, corresponding to a speed-up factor of one.

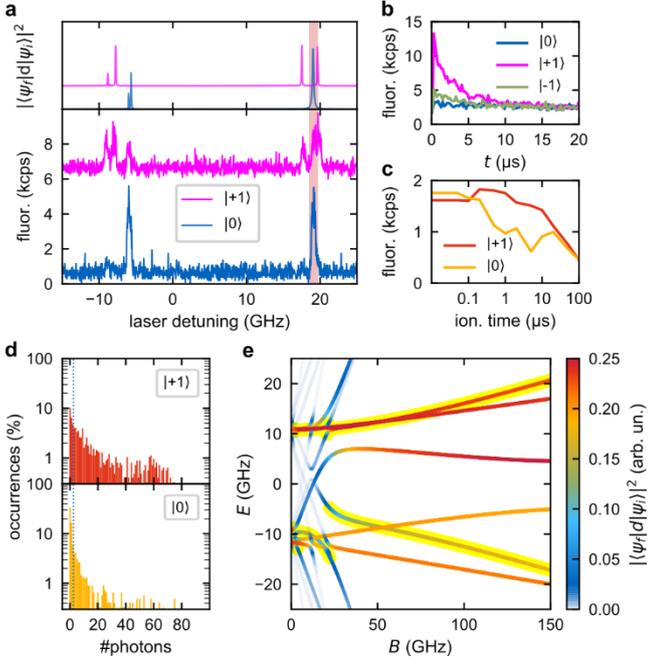

Fig. 4: **Data for a representative shallow implanted NV center (110 keV CN⁻).** **a** PLE spectra as measured (lower panel) and simulated (upper panel) similarly to Fig. 1d. Off-axial strain is estimated to be 12.6 GHz. The detuning is denoted from 637.20 nm. The 637 nm laser was tuned in resonance with the higher-energy transition at 19 GHz for all following measurements (red background). **b** Average fluorescence under optical pumping on the selected transition after explicit spin initialization according to Fig. 1c, as presented in Fig. 2a in the lower panel for the deep NV center. The NV⁻ center was initialized to $|+1\rangle$ with $(70\pm1)\%$ probability. The background after 20 μs stems from finite excitation of the close-by $|\pm1\rangle$ transition. **c** Similar to Fig. 3a. Average fluorescence of the charge-state readout after charge initialization (without postselection), spin initialization in $|+1\rangle$ or $|0\rangle$ according to Fig. 1c, and spin dependent ionization with varying ionization time. For part d, an ionization time of 5 μs was used. **d** Similar to Fig. 3b. Distribution of the number of fluorescence photons for 5 ms of readout. The as-measured fidelity is $(67.1\pm0.9)\%$. **e** Simulation of the dipole transitions between optical ground and excited state for all three spin projection numbers for fixed strain-splitting of 20GHz and varying magnetic field 20° off-axis. The transitions highlighted in yellow are the most pronounced spin $|0\rangle$ transitions. The coloring is a measure for the transition strength. There are several parameter ranges with spectrally well isolated transitions.

**Demonstration on a shallow implanted NV center.** Our method is applicable to "shallow" NV centers less than 100 nm close to the diamond surface. Fig. 4a-d present data recorded on a ~70 nm deep center (110 keV CN⁻ implant; [15]). Spectral lines are inhomogeneously broadened to $(0.43\pm0.02)$ GHz due to spectral diffusion (see Supplementary Section S.10). As the transitions are also much weaker – yielding lower fluorescence – we compensate for this by increasing the resonant red laser power to 240 nW, which is just low enough to prevent pronounced effects by also exciting the close-by $|1\rangle$ transition. Still, this close-by transition compromises spin initialization fidelity (Fig. 4b). Without postselecting on the charge state, the useable contrast was best for 5 μs ionization time and yielded 0.6 kcps and 1.5 kcps for spin $|0\rangle$ and $|+1\rangle$, respectively (Fig. 4c). Including postselection on the charge state, the end-to-end single-shot fidelity, as measured in Fig. 4d, is $(67.1\pm0.9)\%$. Correcting for the non-perfect spin initialization and π-pulse results in a fidelity of $(78.6\pm2.5)\%$, corresponding to a single-shot SNR of 0.99±0.13.

We note that sub-GHz optical linewidths have been reported for comparable implanted NV centers as shallow as 10 nm[16,17] The protocol also promises to be resilient to high-strain environments, since it does not make use of spin-selective intersystem crossing into the $^1A$ singlet state. It only requires a well-separated spin-selective transition. Simulations of the optical transitions according to Doherty *et al.*[10] result in strong transitions separated in the GHz range for high strain upon application of an appropriate magnetic field. In particular using an aligned magnetic field is very powerful. Also for misaligned magnetic fields (and finite strain) well-separated transitions can be identified (Fig. 4e). In turn, we would estimate 1-2 GHz broad lines as the upper limit. According to Fu *et al.*[9] this corresponds to around 35 K. Note that we expect the protocol to work with poorly cycling transitions by using a stronger ionization laser.

Table 1: Overview of the initialization and readout metrics for both NV centers presented in this letter.

|  | deep natural NV center | shallow implanted NV center |
|---|---|---|
| NV⁻ fraction (%) | 99.7±0.7 | 92.9±1.3 |
| spin init. $|+1\rangle$ fraction (%) | 87.9±1.7 | 70.0±1.9 |
| MW error (%) | 5.6±0.1 | 5.1±0.6 |
| end-to-end fidelity (%) | 88.5±0.5 | 67.1±0.9 |
| readout fidelity (%) | 96.4±2.2 | 78.6±2.5 |

## Discussion

We have pushed the fidelity of spin-to-charge conversion into the single-shot regime, by combining it with resonant excitation at cryogenic temperature. The resulting protocol can operate even on shallow implanted NV centers and eliminates the need for any optimized collection optics. We achieve a single-shot SNR of 3.5 and 0.99 on a deep and shallow center respectively, which provides a speedup in the range of $10^3$ over standard readout and a speedup of ~20 over the resonant excitation readout method[7]. As its most important consequence, this technique will enable sensing experiments using long (ms) protocols. These are within the coherence time of shallow NV centers[18], but are currently precluded by acquisition speed. 1 ms of sensing time would enable coherent coupling to a single electron spin at 50 nm distance. In sensing, this would cover the entire thickness of a biological cryoslice[19], in computing it could enable coupling in scalable arrays of NV centers[20]. The protocol is compatible with electric readout of the NV$^-$ spin state[21]; in combination with a single-electron transistor[22], single-shot electric readout might be possible. Our method could also enable single-shot readout of more challenging spin qubits, in particular in Silicon Carbide, where poor photon count rates currently hamper work for some centers with otherwise promising spin properties[23–25].

## Methods

**Experimental setup.** The measurements were performed in a home-built confocal microscope, with the sample being in a flow cryostat using liquid Helium. Inside the vacuum chamber are a movable permanent magnet and a movable air objective (Nikon Plan Apo 40x NA0.95) to illuminate the sample and to collect fluorescence. The fluorescence was separated from the laser illumination with a 650 nm longpass dichroic mirror. Residual laser light was removed with a 650 nm longpass filter and a shortpass filters (800 nm; just relevant for the ionization with NIR, see SI). Photons were detected with an avalanche photo diode (APD). Three individually gated lasers are combined to a single excitation path, so that the sample can be illuminated simultaneously by all of them: A green 517 nm fiber-pigtailed laser diode driven by a PicoLAS LDP-V 03-100 UF3, "cleaned-up" with a 540nm shortpass and combined to the common laser path with a 550 nm longpass dichroic mirror; a red 642 nm fiber-pigtailed laser diode driven by an iC Haus iC-NZN and combined to the external cavity laser's path with a non-polarizing 90:10 beam splitter; a red external cavity diode laser (New Focus TLB-6704) stabilized with a wavemeter (absolute accuracy ±600 MHz) and gated by two AOMs in series. All laser beams were expanded to ~10 mm, which is approx. the back aperture of the used objective. For each beam, we can control the lateral alignment as well as the collimation. Two FPGAs were used to control all short-timescale pulses and to register the APD events.

**Samples preparation.** The deep NV center and the shallow implanted NV center are in two different pieces of diamond. Both diamonds are electronic grade from Element6 and have some spots where CN$^-$ molecules were implanted. Before implanting both diamonds were cleaned with a 1:1:1 mixture of sulfuric:nitric:perchloric acid. Afterwards, they were annealed at 900°C for 3h and cleaned again with the 3-acid mixture. The diamond with the shallow implanted NV presented in the main text was additionally annealed for a second time at 1200°C for 2h. Before measurements, both diamonds were treated in an Oxygen plasma. To remove high fluorescence in the surrounding of shallow NV centers, we illuminate the region with a high-power (~50 mW) green 532 nm laser after cooling down.

## Data availability

The data that support the findings of this study are available from the corresponding author on reasonable request.

## Acknowledgements

We thank Marcus Doherty for helpful discussion.
This work has received support from the Deutsche Forschungsgemeinschaft (DFG) under grants RE3606/1-1, RE3606/2-1, RE3606/3-1 and EXC-2111 – 390814868, from the European Union's Horizon 2020 research and innovation programme under grant agreement No 820394 (ASTERIQS) and from the National Natural Science Foundation of China (NSFC) under grants 11761131011, 81788101, 91636217, 11722544. D.K. was supported by funds of Polish National Science Center (NCN), PhD Student Scholarship No. 2018/28/T/ST3/00390 and Grant No. 2015/19/B/ST3/03152.

**Author contributions**
D.M.I and F.R. designed the experiment. D.M.I. built the setup. D.M.I., F.P. and F.K. conducted the experiment. D.M.I. and F.P. analyzed the data with support from F.R., F.P., F.K., F.S. and J.D. D.M.I. developed all numerical models with support from D.K. and F.P. M.K.. T.L. and J.M. prepared the sample. D.M.I., F.P. and F.R. wrote the manuscript. All authors commented on the manuscript.

**Competing Interests**
The authors declare no competing interests.

**Additional Information**
Supplementary information is available for this paper online.
Correspondence and requests for materials should be addressed to F.R.

# Supplementary Information for
# "Robust All-Optical Single-Shot Readout of Nitrogen-Vacancy Centers in Diamond"

## S.1 Hyperfine Structure and Magnetic Field Evaluation

Standard pulsed Optically Detected Magnetic Resonance (ODMR) techniques were employed for the evaluation of bias magnetic field strength and amplitude. By reducing the MW power used to excite the NV$^-$ spin with a $\pi$-pulse, it is possible to selectively address the hyperfine energy levels resulting from the interaction between the NV$^-$ electron spin and the $^{14}$N nuclear spin $I = 1$ that forms the color center itself. The overall ground state Hamiltonian $\hat{\mathcal{H}}_g$ is:

$$\hat{\mathcal{H}}_g = \hbar \left( D_g \hat{S}_z^2 + \gamma_e \hat{S} \cdot \mathbf{B} + \hat{S} \hat{A} \hat{I} + Q_g \hat{I}_z^2 + \gamma_n \hat{I} \cdot \mathbf{B} \right).$$

The first term in the above Hamiltonian is dominant and can represent the system energy in the absence of external fields, i.e., is the zero-field term, where $\hbar$ is the reduced Planck constant. $\hat{S}$ is the electronic spin operator, and $D_g \simeq 2.87\,\text{GHz}$ the axial zero-field parameter. In the following terms, $\mathbf{B}$ is the external bias magnetic field, $\gamma_e \simeq 2\pi \times 28\,\text{MHz/mT}$ and $\gamma_n \simeq 2\pi \times -3.08\,\text{kHz/mT}$ the electronic and nuclear gyromagnetic ratio, respectively, $Q = -4.945\,\text{MHz}$ the nuclear quadrupole interaction, and $\hat{A}$ the hyperfine spin tensor with axial $A_\parallel = -2.16\,\text{MHz}$ and orthogonal $A_\perp = -2.62\,\text{MHz}$ components.

By solving the eigenvalue equation and considering the hyperfine transition energies as solution of the system, it is possible to determine the magnetic field amplitude $B$ and orientation $\theta$ with respect to the NV quantization axis, the only unknown parameters. In this way, we measured $B = (0.7 \pm 0.1)\,\text{mT}$ and $\theta = (39 \pm 7)°$ for the deep NV center used in the main text. We want to stress that this was a non-optimized situation due to technical limitations and a first indication that the field alignment requirements are not strict. For the shallow NV center also studied in the main text, pulsed ODMR measurements were not able to resolve a more complex hyperfine structure, which highlight interactions with multiple nuclear spins related to other $^{13}$C or N impurities. Nevertheless, since both the rough magnet position and the $|0\rangle \to |\pm 1\rangle$ transitions frequency range correspond to the case related to the deep NV, we can assume $B \leq 1\,\text{mT}$, too.

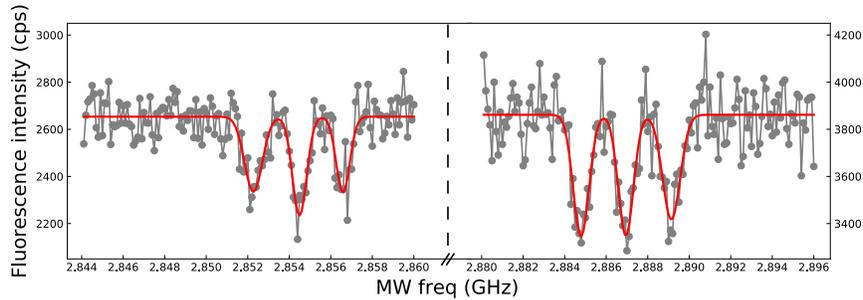

Figure S.1: Hyperfine ODMR transitions measured at room temperature. Grey scatter and red lines are experimental data and Lorentzian fit, respectively.

## S.2 Simulation of the Excited State Structure and Optical Transitions

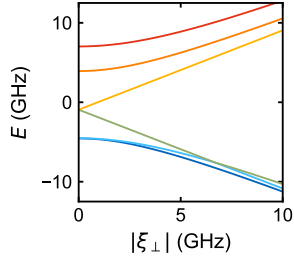

Figure S.2: Excited state structure of an NV$^-$ center at low temperature depending on the non-axial strain. The magnetic flux density is 20 MHz and 45° misaligned. Coloring is for the readers convenience.

We simulated the NV$^-$ excited state structure and the related optical transitions between ground and excited state according to Doherty *et al.*[1].

We solve the Hamiltonian as follows for 14 energy levels:

$$\mathcal{H} = V_{\text{opt}} + V_{\text{ss}} + V_{\text{so}} + \hat{O}_{E,x}(0, \xi_\perp) + \hat{S}_z + \hat{S}_x + \hat{S}_y$$

Here, $V_{\text{opt}}$ is a diagonal matrix with the rough energy differences without considering structure within the ground and excited state. $V_{\text{ss}}$ and $V_{\text{so}}$ are the spin-spin and spin-orbit interaction potential according to Tab. 3 and Tab. 2 in Doherty *et al.*[1]. The values for the entries are taken from Tab. 4 in Doherty *et al.*[1]. $\hat{O}_{E,x}$ is an orbital operator according to Tab. A.4 in Doherty *et al.*[1], where we set the two different entries $O_{a,x} \to \frac{1}{\sqrt{2}} \langle a_1 ||V_E||e\rangle$ to 0 and $O_{b,x} \to \frac{1}{\sqrt{2}} \langle e||V_E||e\rangle$ to the non-axial strain $\xi_\perp$. $\hat{S}_{x,y,z}$ are the components of the total spin operator according to Tab. A.5 in Doherty *et al.*[1]. We set the entries $S_i$ to the component of the magnetic flux density **B** in the "respective" direction. $B_z$ is the component parallel to the NV center axis and determined as half the splitting from an ODMR spectrum.

The transition matrix element is a measure for the transition strength between the inital state **i** and the final state **f**.

$$\mathcal{M}_{fi} = \left| \left\langle \mathbf{f} \left\| \hat{O}_{E,x}(1,0) + \hat{O}_{E,y}(1,0) \right\| \mathbf{i} \right\rangle \right|^2$$

Similarly to $\hat{O}_{E,x}$, $\hat{O}_{E,y}$ is an orbital operator according to Tab. A.4 in Doherty *et al.*[1], where we set the two entries $O_{a,y} \to \frac{1}{\sqrt{2}} \langle a_1||V_E||e\rangle$ to 1 and $O_{a,y} \to \frac{1}{\sqrt{2}} \langle e||V_E||e\rangle$ to 0.

For creating the simulated PLE spectra presented in Fig. 1d and 4a of the main text, we assumed Lorentzian broadening, so that the expected fluorescence is

$$f(E) = \sum_l \frac{A}{\sqrt{\pi\gamma}} \cdot \frac{\gamma^2}{\gamma^2 + (E - E_l)^2}$$

We set the amplitude $A$ to $\mathcal{M}_{fi}$ and the FWHM $\gamma$ to $\mathcal{M}_{fi}/10$. $E_i$ is the energy of the different transitions, i.e. $(E_f - E_i)$.

In literature, the lower-energy spin $|0\rangle$ transition ($E_y$ branch) is reported to have more spin mixing with the $|\pm 1\rangle$ states and is accordingly less cycling.[2,3] This statement is typically made with the exception of low strain, which is the case of the used deep NV center, where we measured the lower-energy transition to be more stable. This is in agreement with the simulation, where this transition is well separated in energy as well as the related excited state level (Fig. S.2 green line).

## S.3   Saturation Curves

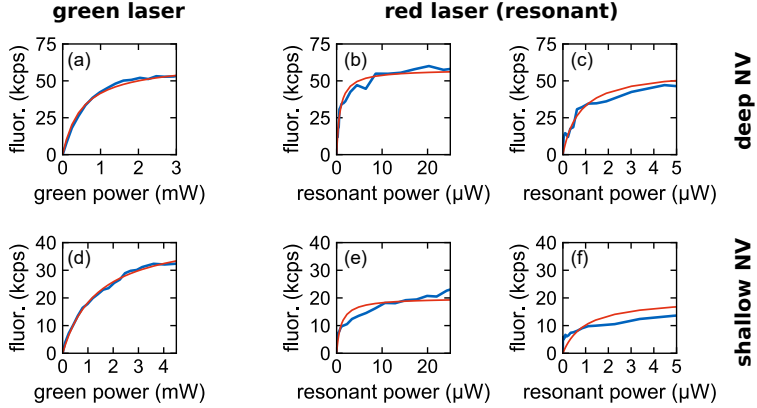

Figure S.3: Saturation curves for both NV centers presented in the main text. Both for illuminating them with a green 517 nm laser (a,d) and a narrow-band laser tuned to an $NV^-$ spin $|0\rangle$ transition (b,e). Panels c and f are zoom-ins to the low-power regime of panels b and e, respectively. The red lines are fits to the data.

Table S.1: Overview of the saturation powers and the related saturation fluorescence for the data presented in Fig. S.3.

|            | $f_{\text{sat,g}}$ (kcps) | $I_{\text{sat,g}}$ (mW) | $f_{\text{sat,r}}$ (kcps) | $I_{\text{sat,r}}$ (µW) |
|---|---|---|---|---|
| deep NV    | $63 \pm 6$  | $0.51 \pm 0.04$ | $58 \pm 10$  | $0.74 \pm 0.09$ |
| shallow NV | $44 \pm 3$  | $1.38 \pm 0.07$ | $20 \pm 333$ | $1 \pm 12$      |

We measured the saturation behavior for both the deep NV center and the shallow NV center presented in the main text. For both NVs, we determined the saturation under illumination with a green 517 nm laser with 20 % duty cycle, where Fig. S.3a,d displays the duty-cycle corrected data. To measure the saturation behavior for the narrow-band red 637 nm laser tuned into resonance with a spin $|0\rangle$ transition (Fig. S.3b,c,e,f), we first initialized the NV centers charge and spin state with the green laser (without postselection). The actual illumination with the red laser is 100 ns short to acquire data without strong depolarization effects.

To determine the saturation power and fluorescence, we fitted the data with $f = A\frac{I \cdot I_{\text{sat}}}{I + I_{\text{sat}}}$.[4,5] $f_{\text{sat}} = A \cdot I_{\text{sat}}$.

For the deep NV center, the saturation count rate under resonant illumination is similar to the saturation value under green illumination. However, for the shallow implanted NV center, it differs a lot, and we cannot observe a clear saturation behavior when illuminating resonantly. We attribute this to spectral diffusion of the shallow implanted $NV^-$ center's optical transition; increasing the laser power leads to power-broadening and in turn the laser line "hits" the optical transition more often. Note that in case of charge instability, increasing the resonant power is expected to shift the charge state balance more and more towards the dark $NV^0$ charge state because 637 nm illumination can cause ionization to $NV^0$ but cannot cause recombination from $NV^0$. Hence, we exclude this effect as a source for the non-saturation behavior; in particular it is even a slight indication that the charge state seems to be reasonably stable in our case.

## S.4  Speed-Up Factor

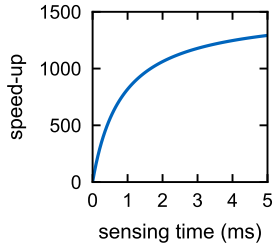

Figure S.4: Estimated speed-up when using the readout protocol presented in this manuscript in comparison with conventional fluorescence-based readout.

To determine the speed-up in comparison to the conventional off-resonant readout with a green laser, we estimated the time necessary to get an (average) SNR larger than one. For the single-shot protocol it is always only one repetition with length of 850 µs plus the time for the actual sensing sequence (e.g. a XY8 sequence). For the conventional readout, the single-shot SNR was calculated as described in Section S.11. The average SNR is $\sqrt{N} \cdot \text{SNR}$, where $N$ is the number of repetitions. Per repetition, we estimated 1.5 µs plus the actual sensing sequence.

Fig. S.4 shows the speed-up as function of the sensing sequence length. We assumed that the conventional readout can take advantage of the full saturation countrate of 50 kcps with a fluorescence contrast of 30 % for 250 ns between the different spin states. Note that the speed-up factor is already 2 for a zero-length sensing sequence.

As the presented protocol relies on low temperature, we also compare it with the resonant excitation readout method presented by Robledo *et al.*[8]. We measured histograms for both their method and our method in a higher-strain environment (∼5.7 GHz) because strain has changed over time. To have the highest comparability possible experimental parameters were as similar as possible by three measures taken. 1) Both measurements were taken within an hour. 2) The resonant excitation readout was performed by just slightly modifying the pulse protocol used to measure our protocol: The ionization pulse was removed, and during the final (charge-state) readout the cw MW excitation was omitted to have a spin-dependent signal as used by Robledo *et al.*[8]. 3) All laser powers were kept to the same value, except for the final resonant excitation readout pulse, which was performed for different powers between 3 nW and 144 nW.

Our protocol resulted in an end-to-end spin readout fidelity of 83.8 % (SNR 1.38) for 1 ms readout duration and 79.4 % (SNR 1.04) for 200 µs readout duration. However, in our setup with a deep NV below a planer surface, the resonant excitation method performed much below an SNR of 1: the fidelity was only 52.8 % (SNR 0.22), best for 100 µs readout at 92 nW. Fig. S.5 presents the related count statistics. In turn, to get an average SNR of 1, 20 repetitions are necessary based on the single-shot SNR stated above.

Considering the same overhead of around 730 µs per repetition (mainly due to 500 µs for postselection)

for both protocols, our low-temperature spin-to-charge conversion promises to be faster by a factor of ∼20 independent of readout duration compared to the current state-of-the-art readout in low-temperature environments – provided low countrate, e.g. because of poor collection efficiency due to not using photonic structures.

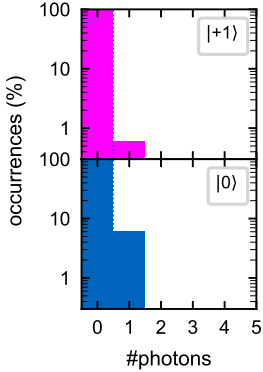

Figure S.5: Count statistics for the resonant excitation readout method according to Robledo *et al.*[8]. The count statistics was taken during 100 µs readout duration with the setup and sample geometry used throughout the rest of this publication; this includes in particular the absence of photonic structures.

## S.5 Spin Initialization

A major limitation concerning both quantum sensing and quantum computing with NV centers is the NV$^-$ spin state initialization. In this section, we present the spin dynamics under resonant optical pumping on an optical transition. First, we exploit the depletion of the related spin state for improving the overall spin initialization. Second, based on these time average measurements, we modeled the spin dynamics with a rate equation that allows to extract the population of each spin state.

### S.5.1 Related Measurements

Commonly in NV center research, the NV center's charge and spin state is initialized into NV$^-$ with spin $|0\rangle$ by illuminating it off-resonantly with a green laser (typically 532 nm). According to Doherty *et al.*[6], "the degree of ground state optical spin- polarisation [into spin $|0\rangle$] is not consistently reported in the literature, with many different values ranging from 42%–96% reported". Hopper *et al.*[7] report a value of around 80 %.

To determine the spin state distribution, we illuminated the NV center with the 637 nm laser at 56 nW tuned into resonance with an optical transition with $|0\rangle$ character. Doing this directly after initialization, the fluorescence intensity correlates with the spin $|0\rangle$ population. To get the actual $|0\rangle$ fraction, the same measurement is necessary for the spin $|\pm 1\rangle$ populations. To access them, we swap the $|0\rangle$ population either with the $|+1\rangle$ or $|-1\rangle$ population by means of a MW $\pi$-pulse. As imperfect $\pi$-pulses can cause additional sources of errors, we also measure the fluoresce after two $\pi$-pulses on the MW $|+1\rangle$ transition. Fig. S.6a shows these four measurements for the deep NV center presented in the main text. The red lines are fits to the curve to determine the actual values, as discussed below. Except for the $\pi$-$\pi$-pulse curve and the fits, these data are presented in Fig. 2a in the main text, too.

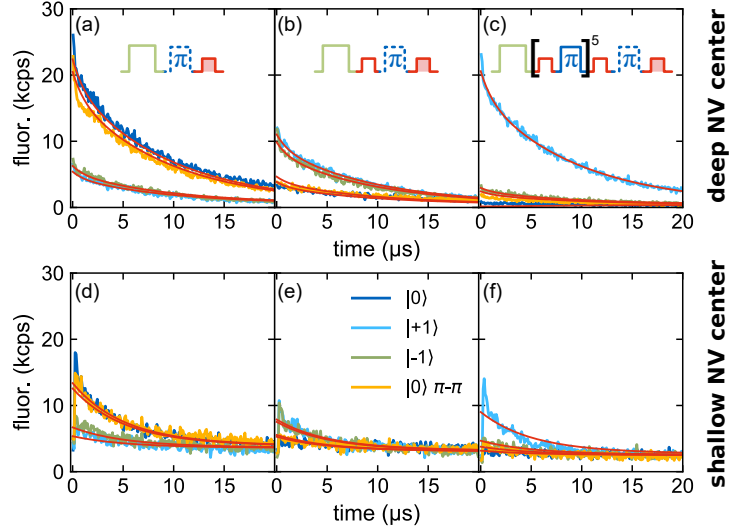

Figure S.6: Fluorescence during optical pumping on a spin $|0\rangle$ transition of the deep NV center (a-c) and the shallow NV center (d-f) used in the main text. Each panel shows four curves; either directly after optical initialization (dark blue), after a MW $\pi$-pulse on the $|+1\rangle$ and $|-1\rangle$ transition (light blue and green) as well as after two $\pi$-pulses on the $|+1\rangle$ transition (orange). In panels a and d, the spin was initialized with a green laser pulse; in b and e, with a green laser pulse followed by a resonant pulse; and in c and f by the full spin initialization protocol. See the insets for the measurement pulse sequences. All traces in each row (a-c; d-f) were measured interleaved. For better distinguishability, the measurement data are averaged over ten bins, with each bin of 10 ns being the average of 500 000 (200 000) experimental repetitions for the deep (shallow) NV center. The red lines are fits to the non-averaged data, see Section S.5.2.

Fig. S.6b displays the spin distribution after 20 µs illumination with the resonant laser; that is the spin distribution after the first measurement without any MW excitation. Similarly to Fig. S.6a, the different spin populations are measured by swapping the population to spin zero by MW $\pi$-pulses. In accordance with Robledo et al.[8], the first pulse has mainly depleted the spin $|0\rangle$ transition and in turn spin $|\pm1\rangle$ is much higher populated.

To exploit this for higher selectivity in spin initialization, after the first resonant pulse we swap the slightly lower-occupied $|-1\rangle$ population back to $|0\rangle$ by a MW $\pi$-pulse. Repeating this several times, more and more population accumulates in the other spin one state $|+1\rangle$, as can be seen in Fig. S.6c.

Fig. S.6d-f present the same measurements for the shallow NV center used in the main text. The pronounced difference is the curves not decaying to zero but to a finite value. This stems from simultaneously also exciting an optical $|\pm1\rangle$ transition, which is spectrally close to the used optical $|0\rangle$ transition (compare Fig. 4a in the main text). This has implications on the used sequence for spin polarization: As longer illumination will mix the spin populations more and more, we shortened each optical pumping pulse to 10 µs. The resonant laser was operated at 170 nW, which is the power, where just a slight decrease in fluorescence happens during resonant illumination together with cw MW spin mixing, as used for the charge-state readout.

## S.5.2 Rate Equation Model

To determine the spin-state distribution and in turn the spin-initialization fidelity, we modeled the NV center dynamics as rate equations with five states. The states are 1) NV$^-$ $|0\rangle$, 2) NV$^-$ $|+1\rangle$, 3) NV$^-$ $|-1\rangle$, 4) NV$^-$ singlet and 5) NV$^0$.

$$n = \begin{pmatrix} n_{|0\rangle} \\ n_{|+1\rangle} \\ n_{|-1\rangle} \\ n_{\text{sing}} \\ n_{\text{NV}^0} \end{pmatrix}$$

$n_i$ is the time-average population of the NV center being in state $i$. At the beginning of each fit, the populations of the three different spin states add to one, while the probabilities to be in the singlet or in NV$^0$ are set to 0.

The dynamics while pumping on an optical transition with spin $|0\rangle$ character is modelled with the transfer matrix as follows:

$$T_{\text{opt}} = \begin{pmatrix} & 0 & 0 & p_{\text{ts}}/2 & 0 \\ 0 & & 0 & p_{\text{ts}}/4 & 0 \\ 0 & 0 & & p_{\text{ts}}/4 & 0 \\ p_{\text{st},0} & p_{\text{st},1} & p_{\text{st},1} & & 0 \\ p_{\text{ion}} & p_{\text{ion}} & p_{\text{ion}} & 0 & \end{pmatrix}$$

To maintain probabilities, the empty diagonal entries are 1 minus the sum of the other elements in the respective column; e.g. the 5$^{\text{th}}$ diagonal element is 1. $p_{\text{st},0}$ and $p_{\text{st},1}$ are the probabilities of having an inter-system crossing to the singlet in one time step, starting from spin $|0\rangle$ or $|\pm 1\rangle$, respectively. $p_{\text{ts}}$ is the probability of an inter-system crossing back to the triplet, where we assumed that $\frac{1}{2}$ of events end up in spin $|0\rangle$ and the other half is equally distributed to both $|\pm 1\rangle$ states. Finally, $p_{\text{ion}}$ is the probability of ionizing the NV$^-$ center to NV$^0$ within one measurement time bin. We do not include recombination from NV$^0$ to NV$^-$ because of the 637 nm laser's energy per photon and its low intensity render this process very unlikely.

The excited state is excluded in the model, too, as it is implicitly described by the fluorescence parameters $f_0$ and $f_1$. These are used as link to fit the measured fluorescence decay curves, where the total fluorescence was modelled as follows for each time step:

$$f(t) = n_{|0\rangle}(t) \cdot f_0 + n_{|+1\rangle}(t) \cdot f_1 + n_{|-1\rangle}(t) \cdot f_1$$

The MW $\pi$-pulses are modeled as follows (here for a $\pi$-pulse on the $|+1\rangle$ transition):

$$T_{\text{MW},+1} = \begin{pmatrix} E_{\text{MW}} & 1 - E_{\text{MW}} & 0 & 0 & 0 \\ 1 - E_{\text{MW}} & E_{\text{MW}} & 0 & 0 & 0 \\ 0 & 0 & 1 & 0 & 0 \\ 0 & 0 & 0 & 1 & 0 \\ 0 & 0 & 0 & 0 & 1 \end{pmatrix}$$

As the overall fluorescence lowers after each optical pumping step (compare Fig. S.6a/d → b/e → c/f), we also include a "fluorLoss" parameter, which accounts for the reduced overall fluorescence in Fig. S.6b,e (fluorLoss1) as well as d,f (fluorLoss6). We attribute this loss of fluorescence mainly to ionization at the very first resonant illumination, when the spin $|0\rangle$ population is high. In turn, there is a high population in the NV$^-$ excited state, which can get ionized comparably easy even by the low-power 637 nm laser.

Fig. S.7 shows all 12 measured fluorescence time traces under optical pumping. Tab. S.2 summarizes the values from fitting all 12 curves simultaneously with the rate equation model described above.

Table S.2: NV$^-$ spin state distribution according to fitting with the rate equation model described in Sec. S.5.2 (upper two sub-tables) and global fitting parameters for all panels in Fig. S.7 together (third sub-table). In the upper two sub-tables, a-f refer to the panels in Fig. S.6 and S.7. The empty cell denote a lifetime that is several orders of magnitude larger, in the range of hours. Lifetimes ($1/e$) are calculated as $T_i = -\text{binwidth}/\ln(1 - p_i)$.

| | deep NV | | |
|---|---|---|---|
| | a | b | c |
| $n_{|0\rangle}$ | $70.4 \pm 1.2$ | $14.8 \pm 0.6$ | $0.0 \pm 1.1$ |
| $n_{|+1\rangle}$ | $13.4 \pm 0.6$ | $45.1 \pm 0.4$ | $87.9 \pm 1.7$ |
| $n_{|-1\rangle}$ | $16.3 \pm 1.3$ | $40.1 \pm 0.7$ | $12.1 \pm 2.0$ |
| $\mathcal{F}_{\text{spin}}$ | $85.2 \pm 0.6$ | $72.6 \pm 0.2$ | $94.0 \pm 0.9$ |

| | shallow implanted NV | | |
|---|---|---|---|
| | d | e | f |
| $n_{|0\rangle}$ | $70.4 \pm 1.8$ | $18.9 \pm 0.9$ | $6.9 \pm 1.4$ |
| $n_{|+1\rangle}$ | $9.7 \pm 1.2$ | $42.1 \pm 0.8$ | $70.0 \pm 1.9$ |
| $n_{|-1\rangle}$ | $19.8 \pm 2.2$ | $38.9 \pm 1.2$ | $23.1 \pm 2.3$ |
| $\mathcal{F}_{\text{spin}}$ | $85.2 \pm 0.9$ | $71.1 \pm 0.4$ | $85.0 \pm 0.9$ |

| | deep NV | shallow NV |
|---|---|---|
| $E_{\text{MW}}$ (%) | $5.6 \pm 0.1$ | $5.1 \pm 0.6$ |
| $f_0$ (kcps) | $31.7 \pm 0.6$ | $17.5 \pm 0.4$ |
| $f_1$ (kcps) | $0.2 \pm 0.3$ | $3.6 \pm 0.2$ |
| $T_{\text{st},0}$ (µs) | $4.1 \pm 0.3$ | $7.0 \pm 0.7$ |
| $T_{\text{st},1}$ (ms) | $0.4 \pm 0.3$ | $1.0 \pm 10.0^{-6}$ |
| $T_{\text{ts}}$ (µs) | $1.33 \pm 0.09$ | $11.3 \pm 2.9$ |
| $T_{\text{ion}}$ (ms) | | $0.2 \pm 0.6$ |
| fluorLoss1 (%) | $20.5 \pm 0.3$ | $15.1 \pm 0.5$ |
| fluorLoss6 (%) | $21.9 \pm 0.3$ | $30.0 \pm 0.5$ |
| $R^2$ | $0.944$ | $0.537$ |

## S.6 Charge Initialization

The NV charge state was initialized with a green 517 nm laser pulse of 2 µs length at 1.4 mW in case of the deep NV center. The initialization was optimized as follows: After an ionization pulse, different initialization powers and times were applied, followed each by a charge readout step. The initialization (without postselection) that maximized the average fluorescence during the readout step was used.

For the shallow NV center, we used the initialization sequence as follows: red 642 nm laser at

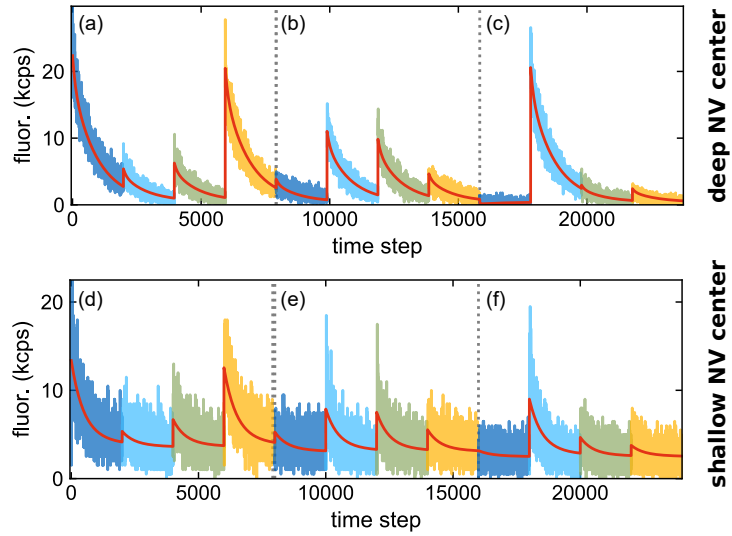

Figure S.7: Fit of the fluorescence decay curves under optical pumping. Same data than in Fig. S.6, including the same color code. Here each measurement bin is shown, as used for the fitting. Each row displays the trace as measured, i.e. the 12 different "sub-curves" were measured as a whole before the next repetition of all "sub-curves". To minimize effects related to the previous sequence, an additional green (2 µs) and resonant (20 µs) pulse was applied, before next the "sub-curve" starts with the respective initialization.

>17 mW for 1 µs → 500 ns break → green 517 nm laser at 3.6 mW for 3 µs → 200 ns break → green 517 nm laser at 3.6 mW for 200 ns. This performed better than a single green initialization pulse with respect to the average fluorescence related to a charge readout afterwards (see above for the optimization procedure). We attribute this to changing the local surrounding of the shallow implanted NV center. Such changes were demonstrated to shift the charge-state balance.[9]

To improve the charge initialization fidelity for the final measurements, we applied a charge readout step of 500 µs (1 ms) directly after the initialization sequence for the deep (shallow implanted) NV center and postselected on events with at least 6 (2) photons, which corresponds to an acceptance rate of about 37 % (22 %) of repetitions.

To quantify the charge initialization and readout, we measured the photon count statistics for a charge readout both directly after the initialization and after a strong ionization pulse in between. This ionization pulse was 5 µs of resonant laser together with the 642 nm laser, followed by 15 µs with additional cw MW to counteract depolarization. In this supplementary information, we present the data taken for long photon acquisition time as this promises the least error due to charge readout and in turn is the best estimate for the charge initialization to $NV^-$. The resulting $NV^-$ and $NV^0$ count statistics is presented in Fig. S.8, which includes fits with Poisson and Gauß distributions. To distinguish between both charge states, we set a threshold that was determined by minimizing the sum of errors for both distributions, i.e. the percentage of events below (above) threshold for the $NV^-$ ($NV^0$) distribution.

For the deep NV center, the $NV^0$ distribution for 5 ms readout duration can be fitted well ($R^2 = 0.99996$) with a Poison distribution centered around $(0.4712 \pm 0.0004)$ photons. The $NV^-$ distribution can be fitted ($R^2 = 0.927$) with the sum of two Gaussians. We attribute the Gaussian nature to spectral diffusion of the deep NV center's optical transitions, so that the spectral overlap between the laser mode and the NV transition varies from repetition to repetition. The NV$^-$ distribution has 0.25 % of events below threshold (5 photons), which is considered as NV$^0$ fraction after the charge initialization.

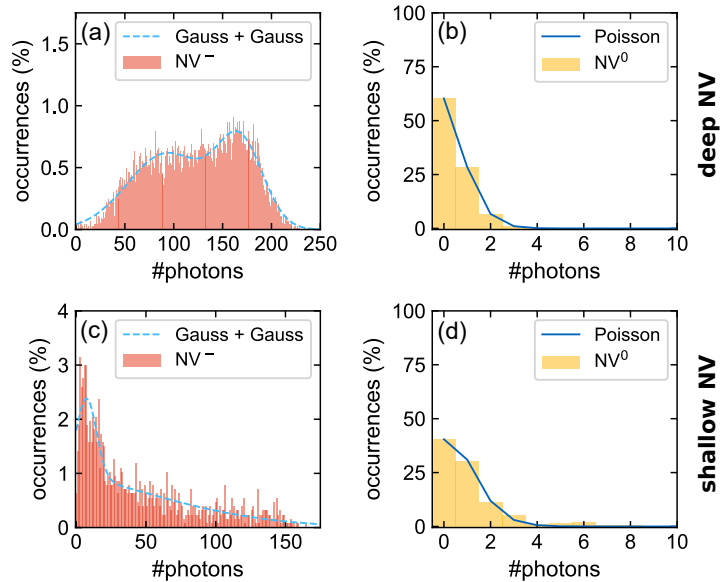

Figure S.8: Count statistics for the charge-state readout. For the deep NV center (a,b) the readout time was 5 ms and for the shallow implanted NV center (c,d) it was 10 ms. The left column displays the distribution after the charge-state initialization, while the right column has a strong ionization pulse included between initialization and readout.

For the shallow NV center, the distributions were evaluated at the maximum readout time of 10 ms. Here, the fraction of the NV$^-$ distribution below threshold (4 photons) is 7.1 %. The NV$^0$ distribution can be fitted well with a Poissonian ($R^2 = 0.995$) around $(0.766 \pm 0.006)$ photons. The NV$^-$ distribution has to be fitted again with the sum of two Gaussians ($R^2 = 0.87$). They are centered around $(-46 \pm 75)$ and $(7.6 \pm 0.6)$ photons with a standard deviation of $(65 \pm 19)$ and $(5.0 \pm 0.6)$ photons. The amplitudes are $(1.1 \pm 0.6)$ and $(1.5 \pm 0.1)$ %.

To determine the threshold for the spin-state distribution, we minimized the charge-state error as described above. The only difference is that we used the same readout time than for the spin-state distribution; compare Fig. 2d and 3b in the main text, which were both acquired with 1 ms of readout time.

## S.7  Ionization with NIR

According to simple energy considerations, the second step for the NV$^-$ ionization might be possible with infrared (IR). The ionization energy for NV$^-$, i.e. the energy difference between the NV$^-$ ground state and the conduction band of diamond, was reported to be 2.60 eV[5]. By just subtracting 1.95 eV (637 nm; zero-phonon line of NV$^-$) from the ionization energy, even a wavelength as long as 1900 nm might serve as second photon for the ionization.

To ionize a deep NV center, we compared using a red 642 nm and an IR 980 nm laser, which both should serve as source for the 2$^{\text{nd}}$ photon. Both lasers are pigtailed laser diodes with single-mode fiber. Fig. S.9 presents measurements similar to Fig. 3a and 4c in the main text. We do see a reduction in fluorescence when illuminating the NV center simultaneously with the resonant laser (which provides the first photon to excite the $NV^-$) and the IR laser (providing the second photon to ionize the $NV^-$). However, compared to the red $642\,\mathrm{nm}$ laser, the ionization time needs to be more than one order of magnitude longer, indicating a much worse absorption cross section for $980\,\mathrm{nm}$. We used the red $642\,\mathrm{nm}$ laser as source for the second photon for all other data presented in this manuscript.

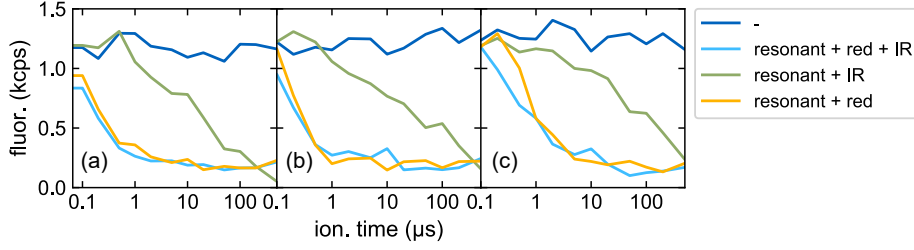

Figure S.9: Fluorescence after ionization with different lasers and powers. The NV center was either subjected to no ionization (dark blue), to a combined pulse of the resonant + red + IR laser (light blue), resonant + IR (green) and resonant + red (orange). The lasers were either (a) at their maximum powers (red at $17\,\mathrm{mW}$ and IR at $33\,\mathrm{mW}$), (b) at $17\,\mathrm{mW}$ and (c) at $10\,\mathrm{mW}$. All traces within each panel were taken interleaved.

## S.8 Correcting the Readout Fidelity

The spin fidelity as measured in Fig. 3b and 4d of the main text is a measure of the end-to-end performance of the whole protocol. This includes the spin-dependent ionization and the readout step itself, but also non-perfect charge and spin initialization as well as errors of the MW. These additional error sources due to initialization and MW are independent of the readout that is chosen and depend on the overall technical implementation. To correct for these effects and to get the actual fidelity related to just our readout scheme, we model the whole protocol as a multi-step process, see Fig. S.10.

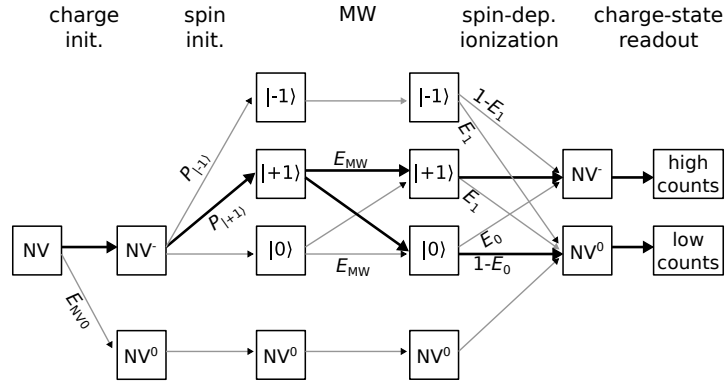

Figure S.10: Model of the whole protocol. Steps are charge and spin initialization, MW $\pi$ pulse (or no MW), spin-dependent ionization and the final charge-state readout.

Here, all variables are probabilities, which means that at any branching, the different leaving paths add up to 1. $E_{\mathrm{NV0}}$ is the population in the neutral $NV^0$ charge state after charge initialization (including postselecting on the charge state). $P_{|-1\rangle}$ and $P_{|+1\rangle}$ are the probabilities of having spin $|-1\rangle$ and $|+1\rangle$, respectively. Accordingly, the probability of initializing into spin $|0\rangle$ is $(1 - P_{|-1\rangle} - P_{|+1\rangle})$. $E_{\text{MW}}$ is the error of a MW $\pi$-pulse, meaning that with this probability, the $\pi$ pulse does not flip the spin. $E_0$ is the error of not ionizing spin $|0\rangle$ despite having tuned the resonant laser to a $|0\rangle$ transition, and $E_1$ the error of ionizing spin $|\pm 1\rangle$. The bold black arrows indicate the paths for perfect experimental conditions. This includes the explicit spin initialization into $|+1\rangle$, which was used for the spin-dependent count statistics presented in Fig. 3b (4d) for the deep (shallow) NV center.

It is important to note, that the actual fidelity of our readout protocol just refers to the steps 'spin-dependent ionization' and 'charge-state readout', which are modeled together with the parameters $E_0$ and $E_1$. On the other side, the errors that are used to determine the as-measured end-to-end fidelity are $E_{0,\text{meas}}$ and $E_{1,\text{meas}}$. $E_{0,\text{meas}}$ ($E_{1,\text{meas}}$) is the sum of all paths that end up in high (low) counts, while intending to prepare the spin in $|0\rangle$ ($|+1\rangle$) by means of a MW $\pi$-pulse (no MW $\pi$-pulse) in step 3 after the spin initialization.

Summing up all paths, we get a linear equation system with two unknown variables $E_0$ and $E_1$.

$$
\begin{aligned}
E_{0,\text{meas}} = \quad & (1 - E_{\text{NV0}}) \cdot P_{|-1\rangle} \cdot (1 - E_1) \\
+ \quad & (1 - E_{\text{NV0}}) \cdot P_{|+1\rangle} \cdot E_{\text{MW}} \cdot (1 - E_1) \\
+ \quad & (1 - E_{\text{NV0}}) \cdot P_{|+1\rangle} \cdot (1 - E_{\text{MW}}) \cdot E_0 \\
+ \quad & (1 - E_{\text{NV0}}) \cdot (1 - P_{|-1\rangle} - P_{|+1\rangle}) \cdot (1 - E_{\text{MW}}) \cdot (1 - E_1) \\
+ \quad & (1 - E_{\text{NV0}}) \cdot (1 - P_{|-1\rangle} - P_{|+1\rangle}) \cdot E_{\text{MW}} \cdot E_0 \\
\\
E_{1,\text{meas}} = \quad & (1 - E_{\text{NV0}}) \cdot P_{|-1\rangle} \cdot E_1 \\
+ \quad & (1 - E_{\text{NV0}}) \cdot P_{|+1\rangle} \cdot E_1 \\
+ \quad & (1 - E_{\text{NV0}}) \cdot (1 - P_{|-1\rangle} - P_{|+1\rangle}) \cdot (1 - E_0) \\
+ \quad & E_{\text{NV0}}
\end{aligned}
$$

The variables and solutions for the deep and shallow NV center are summarized in Tab. S.3.

Table S.3: Overview of the measured end-to-end spin fidelity $\mathcal{F}_{\text{meas}}$ as well as the probabilities/errors that were estimated for the initialization with the explicit spin initialization and for the MW. These values were taken into account when solving the linear equation system with solutions $E_0$ and $E_1$. The charge initialization was assumed to be perfect. Photons were collected for 1 ms (10 ms) for the deep (shallow) NV center.

|  | deep NV | shallow NV |
|---:|:---:|:---:|
| $E_{0,\text{meas}}$ (%) | $17.6 \pm 0.7$ | $44.3 \pm 1.3$ |
| $E_{1,\text{meas}}$ (%) | $5.4 \pm 0.7$ | $21.4 \pm 1.3$ |
| $\mathcal{F}_{\text{meas}}$ (%) | $88.5 \pm 0.5$ | $67.1 \pm 0.9$ |
| $P_{|-1\rangle}$ (%) | $12.1 \pm 2.0$ | $23.1 \pm 2.3$ |
| $P_{|+1\rangle}$ (%) | $87.9 \pm 1.7$ | $70.0 \pm 1.9$ |
| $E_{\text{MW}}$ (%) | $5.6 \pm 0.1$ | $5.1 \pm 0.6$ |
| $E_0$ (%) | $1.8 \pm 2.2$ | $25.3 \pm 3.1$ |
| $E_1$ (%) | $5.4 \pm 2.5$ | $17.4 \pm 2.5$ |
| $\mathcal{F}$ (%) | $96.4 \pm 2.2$ | $78.6 \pm 2.5$ |
| single-shot SNR | $3.5 \pm 1.2$ | $0.99 \pm 0.13$ |

## S.9 Photon Collection Statistics for Short Readout

Table S.4: Readout metrics for different readout times. End-to-end charge and spin fidelities of the deep NV center presented in the main text, as well as fidelities corrected by spin-initialization and MW imperfections. The data for 1 ms refer to the data presented in the main text. Note that the charge count statistics are not taken into account for correcting the spin fidelity—it is presented here to have a more clear estimate on the influence of the charge readout fidelity.

|  | 10 ms | 1 ms | 100 µs | 50 µs |
|---:|:---:|:---:|:---:|:---:|
| $E_{\text{NV0}}$ w/o post-sel. (%) | $46.1 \pm 0.7$ | $49.0 \pm 0.7$ | $60.0 \pm 0.7$ | $70.9 \pm 0.7$ |
| $E_{\text{NV0}}$ (%) | $0.3 \pm 0.7$ | $0.4 \pm 0.7$ | $14.9 \pm 0.7$ | $35.0 \pm 0.7$ |
| threshold (photons) | 6 | 3 | 1 | 1 |
| $\mathcal{F}_{\text{charge,meas}}$ (%) | $98.2 \pm 0.5$ | $98.1 \pm 0.5$ | $85.3 \pm 0.5$ | $76.8 \pm 0.5$ |
| $E_{0,\text{meas}}$ (%) | $17.7 \pm 0.7$ | $17.6 \pm 0.7$ | $16.2 \pm 0.7$ | $12.2 \pm 0.7$ |
| $E_{1,\text{meas}}$ (%) | $5.2 \pm 0.7$ | $5.4 \pm 0.7$ | $19.2 \pm 0.7$ | $37.3 \pm 0.7$ |
| $\mathcal{F}_{\text{meas}}$ (%) | $88.5 \pm 0.5$ | $88.5 \pm 0.5$ | $82.3 \pm 0.5$ | $75.2 \pm 0.5$ |
| $E_0$ (%) | $1.9 \pm 2.2$ | $1.8 \pm 2.2$ | $2.9 \pm 1.9$ | $1.8 \pm 1.6$ |
| $E_1$ (%) | $5.2 \pm 2.5$ | $5.4 \pm 2.5$ | $19.2 \pm 2.2$ | $37.7 \pm 1.7$ |
| $\mathcal{F}$ (%) | $96.5 \pm 0.2$ | $96.4 \pm 2.2$ | $88.9 \pm 1.8$ | $80.4 \pm 1.5$ |
| single-shot SNR | $3.6 \pm 1.2$ | $3.5 \pm 1.2$ | $1.8 \pm 0.2$ | $1.21 \pm 0.10$ |

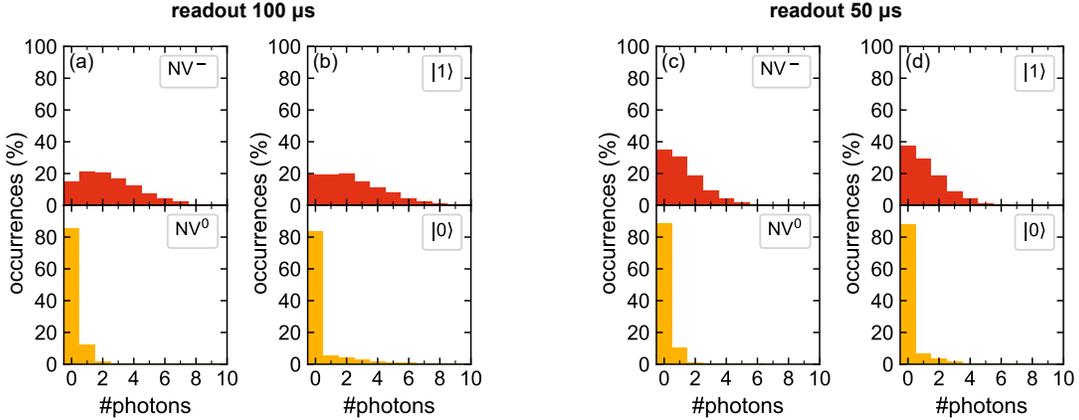

Figure S.11: Count statistics for 100 µs and 50 µs readout duration with the deep NV center presented in the main text. Parts a,c (b,d) were measured simultaneously with the data presented in Fig. 2d (3b) in the main text.

The charge and spin count statistics (histograms) of the readout presented in Fig. 2d and 3b were measured with an actual readout duration of 10 ms. However, the software measured count statistics not only for the whole 10 ms, but also after the first 0.05, 0.1, 0.2, 0.5, 1, 2 and 5 ms. (In the main text, they were evaluated for 1 ms.) In the first place, this allows to compare different readout times for otherwise exact same measurement parameters. In particular, it indicates that for the already quite low saturation count rate of 50 kcps for this NV center in this setup we can speed up the measurement a lot by reducing the readout time to 100 µs, while still maintaining an end-to-end fidelity of $> 79\,\%$, which corresponds to a single-shot SNR $> 1$[7]. Tab. S.4 summarizes the end-to-end fidelities for readout durations spanning two orders of magnitude and Fig. S.11 presents some related count statistics.

In the second place, this simultaneous acquisition of count statistics for different readout times allows to mimic the situation for poor collection optics: Evaluating just the first 100 µs of readout still gives a good directly measured end-to-end fidelity of 82.3 %. Together with the fluorescence being stable over the whole 10 ms of laser illumination, it resembles the situation of $1/100^{\text{th}}$ the detected fluorescence count rate and taking the count statistics for the whole 10 ms. With the NV center used for taking these data having a saturation count rate of 50 kcps, a single-shot readout would be possible even for a saturation count rate as low as 500 clicks per second.

## S.10 Spectral Diffusion

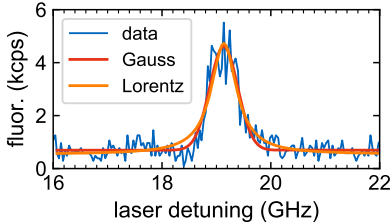

Figure S.12: Fitting the broadened optical transition. Section of the PLE spectrum presented in Fig. 4a in the main text. The used optical transition was fitted both with a Gaussian and Lorentzian curve in this section.

To determine to which extent the PLE linewidth of the shallow implanted NV used in the main text is limited by spectral diffusion, we fitted the used $|0\rangle$ transition both with a Gaussian and a Lorentzian curve (Fig. S.12). While the Lorentzian fit has an $R^2$ of 0.69, the Gaussian fit performs better with an $R^2$ of 0.87. This indicates that inhomogeneous broadening due to spectral diffusion is the main reason for broadening, with a Gaussian FWHM of $(0.43 \pm 0.02)$ GHz. Note that the initialization procedure for this measurement was different with green illumination for 500 ns at 1.2 mW. In contrast, the initialization in the other measurements with the shallow NV center include a longer (3 µs) and stronger (3.6 mW) green pulse and in addition a >17 mW red pulse for 1 µs. Thus, the inhomogeneously broadened linewidth in the relevant measurements is expected to be even broader.

## S.11 SNR and Fidelity Calculation

The numbers for the single-shot SNR and the fidelity are calculated according to Hopper *et al.*[7]. For the common off-resonant readout with a green laser, the single-shot SNR is estimated as

$$\text{SNR} = \frac{\#ph - 0.7 \cdot \#ph}{\sqrt{\#ph + 0.7 \cdot \#ph}}$$

where $\#ph = f_{\text{sat}} \cdot 250\,\text{ns}$ is the photon number per readout repetition and $f_{\text{sat}}$ is the saturation count rate when illuminating the NV center with a green laser. We assume the contrast between spin states to be 0.3.

For our protocol, we use thresholding to separate between the charge states and in turn the spin states. The single-shot SNR was estimated as

$$\text{SNR} = \frac{1 - E_1 - E_0}{\sqrt{(1 - E_1) \cdot E_1 + (1 - E_0) \cdot E_0}}.$$